\documentclass[footinbib,fleqn,aps,prl,twocolumn,superscriptaddress,reprint]{revtex4-2}

\usepackage[]{amsmath}
\usepackage{amssymb}
\usepackage[dvips]{graphicx}
\usepackage{color}
\usepackage{tabularx}
\usepackage{mathtools}
\usepackage{algpseudocode}
\usepackage{enumitem}
\usepackage{multirow}
\usepackage{epstopdf}  
\usepackage{bm}

\usepackage{makecell}
\usepackage{hyperref}
\usepackage[version=3]{mhchem}
\usepackage{tikz}
\usetikzlibrary{shapes,snakes}
\usetikzlibrary{arrows}
\usepackage{threeparttable}

\begin{document}

\title{Data-Driven Prediction of Complex Crystal Structures of Dense Lithium}

\author{Xiaoyang Wang}
\affiliation{State Key Laboratory of Superhard Materials \& International Center for Computational Method and Software, College of Physics, Jilin University, Changchun 130012, P.R.~China}
\affiliation{Laboratory of Computational Physics,
  Institute of Applied Physics and Computational Mathematics, Fenghao East Road 2, Beijing 100094, P.R.~China}
  
\author{Zhenyu Wang}
\affiliation{State Key Laboratory of Superhard Materials \& International Center for Computational Method and Software, College of Physics, Jilin University, Changchun 130012, P.R.~China}

\author{Pengyue Gao}
\affiliation{State Key Laboratory of Superhard Materials \& International Center for Computational Method and Software, College of Physics, Jilin University, Changchun 130012, P.R.~China}

\author{Chengqian Zhang}
\affiliation{DP Technology, Beijing 100080, P.R.~China}
\affiliation{College of Engineering, Peking University, Beijing 100871, P.R.~China}

\author{Jian Lv}
\email{lvjian@jlu.edu.cn}
\affiliation{State Key Laboratory of Superhard Materials \& International Center for Computational Method and Software, College of Physics, Jilin University, Changchun 130012, P.R.~China}
  
\author{Han Wang}
\email{wang\textunderscore han@iapcm.ac.cn}
\affiliation{Laboratory of Computational Physics,
  Institute of Applied Physics and Computational Mathematics, Fenghao East Road 2, Beijing 100094, P.R.~China}
\affiliation{HEDPS, CAPT, College of Engineering, Peking University, Beijing 100871, P.R.~China}  
  
\author{Haifeng Liu}
\affiliation{Laboratory of Computational Physics,
  Institute of Applied Physics and Computational Mathematics, Fenghao East Road 2, Beijing 100094, P.R.~China}
    
\author{Yanchao Wang}
\affiliation{State Key Laboratory of Superhard Materials \& International Center for Computational Method and Software, College of Physics, Jilin University, Changchun 130012, P.R.~China}

\author{Yanming Ma}
\email{mym@jlu.edu.cn}
\affiliation{State Key Laboratory of Superhard Materials \& International Center for Computational Method and Software, College of Physics, Jilin University, Changchun 130012, P.R.~China}

	\date{\today}
	
\begin{abstract}
Lithium (Li) is a prototypical simple metal at ambient conditions, but exhibits remarkable changes in structural and electronic properties under compression. There has been intense debate about the structure of dense Li, and recent experiments offered fresh evidence for new yet undetermined crystalline phases near the enigmatic melting minimum region in the pressure-temperature phase diagram of Li. Here, we report on an extensive exploration of the energy landscape of Li using an advanced crystal structure search method combined with a machine learning approach, which greatly expands the scale of structure search, leading to the prediction of four complex Li crystal phases containing up to 192 atoms in the unit cell that are energetically competitive with known Li structures. These findings provide a viable solution to the newly observed yet unidentified crystalline phases of Li, and showcase the predictive power of the global structure search method for discovering complex crystal structures in conjunction with accurate machine-learning potentials.
\end{abstract}
\maketitle

Light alkali metals lithium (Li) and sodium (Na) adopt high-symmetry cubic crystal structure and exhibit nearly free-electron behaviors at ambient conditions. These prototypical simple metals, however, undergo drastic property changes under strong compression, showcasing enhanced superconducting critical temperature $T_c$ \cite{Shimizu2002, Struzhkin.2002.10.1126/science.1078535}, metal–semiconductor/insulator transitions \cite{Ma2009,Matsuoka2009,Lv2011,Marques2011}, anomalous melting curves \cite{Gregoryanz2005,Guillaume2011,Schaeffer2012,Frost2019} and emergence of symmetry-breaking structures \cite{Hanfland2000,Gregoryanz2008}. These remarkable properties are accompanied by a series of transitions into complex crystal structures in the pressure-temperature (P-T) phase diagram, which present formidable challenges to both experimental measurements and theoretical elucidation. The ground state of Li was recently found to adopt an $fcc$ structure rather than the previously recognized $9R$ structure \cite{Ackland2017}. Meanwhile, temperature-induced phase transitions around the melting minimum at high pressures were observed in Na \cite{Gregoryanz2008} and recently proposed for Li \cite{Frost2019,Wang2022}, but determination of pertinent Li structures remains an open question.

Concerted experimental and theoretical efforts in past decades have led to the construction of phase diagram of Li up to $\sim$130 GPa at low temperatures. Its crystal structure transforms from the $bcc$ (298 K) to $cI16$ structure at 42 GPa through the intermediate $fcc$ and $hR1$ structures \cite{Hanfland2000}, followed by a complex phase transition sequence, $cI16 \stackrel{\sim 62 GPa}{\rightarrow}  oC88 \stackrel{\sim 70 GPa}{\rightarrow} oC40 \stackrel {\sim 95 GPa} {\rightarrow} oC24$, accompanied by an intriguing metal-semiconductor-metal transition \cite{Guillaume2011,Marques2011,Lv2011}.
The melting curve of Li reaches a maximum followed by a pronounced minimum \cite{Guillaume2011,Schaeffer2012,Frost2019}, similar to those observed in other alkali metals \cite{Gregoryanz2005,Luedemann.1968.10.1029/jb073i008p02795,McBride.2015.10.1103/physrevb.91.144111,Robinson.2019.10.1073/pnas.1900985116}. The melting minimum of Li occurs in the pressure range of $40-60$ GPa, but the melting temperature ($T_m$) is sensitive to the pre-melting crystal structure and nuclear quantum effects (NQEs). Single crystal X-ray diffraction measurements by Guillaume \textit{et al.} determined $T_m$ to be 190 K \cite{Guillaume2011}. But later resistivity measurements by Schaeffer \textit{et al.} found a higher $T_m$ of 306 K \cite{Schaeffer2012}. The discrepancy was attributed to the emergence of an amorphous structure before melting. However, recent X-ray diffraction measurements with a rapid compression scheme by Frost \textit{et al.} revisited this disputed region and found $T_m$ between 275 and 320 K, below which evidence for new crystalline phases was found \cite{Frost2019}.

{\it Ab initio} simulations \cite{Tamblyn2008,Hernandez2010,Feng2015} predicted $T_m$s between 250-300 K, in fair agreement with experimental measurements \cite{Schaeffer2012,Frost2019}. Calculations using the Wigner-Kirkwood approximation for NQEs in the liquid and the lattice entropy in the solid \cite{Elatresh2016} produced a $T_m$ of 200 K between 40 and 60 GPa, close to the measured value reported in Ref. \cite{Guillaume2011}. All these works regard $cI16$ as the solid phase before melting, and a recent computational study proposed a temperature-induced phase transition from the $cI16$ to a $C2/m$ structure before melting, predicting a $T_m$ of at least 300 K and a pre-melting regime with collective atomic motions \cite{Wang2022}. It is known that Na has crystal structures containing up to 512 atoms in a unit cell near its melting minimum \cite{Gregoryanz2008}. Given the similarities between Li and Na, along with recent computational and experimental findings \cite{Schaeffer2012,Frost2019,Wang2022}, it is reasonable to expect a rich polymorphism and the emergence of complex crystal structures for Li near its melting minimum.

\begin{figure*}
	\begin{center}
		\includegraphics[width=1.00\textwidth]{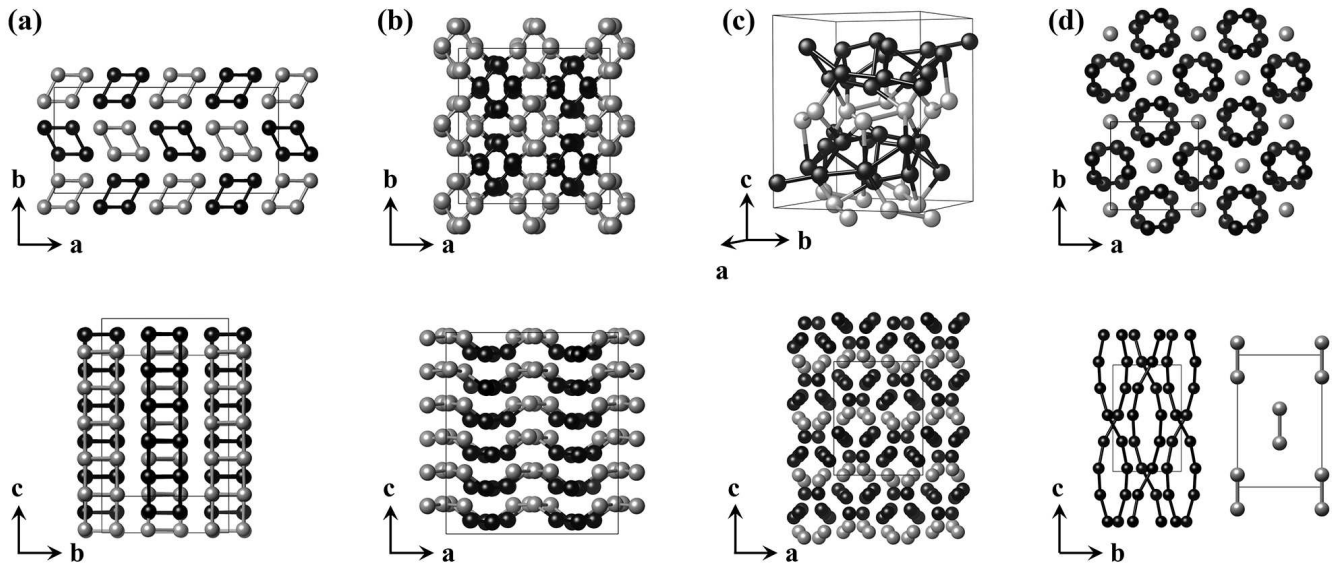}\\[5pt]  
		\caption{Predicted crystal structures of Li. (a) $mP160$, (b) $oP192$, (c) $oP48$, and (d) $tI20$. Two views are given in each case. Atoms in different rhombic prisms in (a), (b) or different layers in (c) are shown in black and grey, respectively, while atoms in the host-guest structures in (d) are shown in black and grey, respectively. See text for detailed descriptions.}
		\label{fig:Fig1}
	\end{center}
\end{figure*}

In this work, we explore the potential energy surface (PES) of Li around the melting minimum using the swarm-intelligence-based CALYPSO method \cite{Wang2010,Wang2012,Lv2012,Gao2019} combined with a machine learning potential named Deep Potential (DP) \cite{Zhang2018,Han2018}. This approach allows examination of crystal structures containing up to 200 atoms, which was previously prohibited by the high computational cost of  global structure searches based on density functional theory (DFT). Our study identifies four crystal structures denoted by Pearson notations of $mP160$, $oP192$, $oP48$ and $tI20$. Among them, the $mP160$ and $oP192$ structures are of the same kind comprising parallel atomic chains, the $oP48$ structure is composed of symmetry-related atomic layers analogous to the experimentally identified $oC88$ and $oC40$ phases, and the $tI20$ structures can be seen as a commensurate host-guess structure corresponding to incommensurate composite structures in heavier alkali metals. 
Gibbs free energy calculations for various experimentally observed and theoretically predicted Li phases indicate a complex energy landscape with multiple shallow minima, suggesting that the four predicted structures are experimentally accessible as promising candidates for the undetermined phases observed in recent experiments \cite{Frost2019}. The present work demonstrates an efficient approach for exploring multi-minima PES and identifying complex crystal structures under diverse P-T conditions, laying the foundation for elucidating broad range of properties.  

Our structure searches used the DP model as the PES calculator \cite{Zhang2018,Han2018} for the CALYPSO method \cite{Wang2010,Wang2012,Lv2012,Gao2019}, which has successfully predicted the structures of a large number of systems \cite{Wang2022Crystal}, including the semiconducting $oC40$ phase of Li \cite{Lv2011}. Taking advantage of this accelerated structure optimization scheme supported by the efficient DP model, we performed high-throughput searches with simulation cells containing 1-200 atoms at 50 and 60 GPa. More than 600,000 structures were sampled, among which about 5,000 lowest-enthalpy structures were further subjected to duplicate elimination and refinement by the DFT calculations. Further computational details for DFT calculations and DP constructions are provided in Supplemental Material \cite{Note1} Sec.~SI and SII, respectively.
Using this approach, we reproduced the experimentally observed $cI16$ and $oC88$ phases of Li and found four additional energetically competitive structures denoted by $mP160$, $oP192$, $oP48$ and $tI20$ shown in Fig.~\ref{fig:Fig1}. These structures contain large numbers of atoms in the unit cell and complex bonding arrangements. Pertinent structure files along with those of other low-lying crystal structures are provided in Supplemental Material \cite{Note1}.

\begin{figure}
\begin{center}
\includegraphics[width=0.5\textwidth]{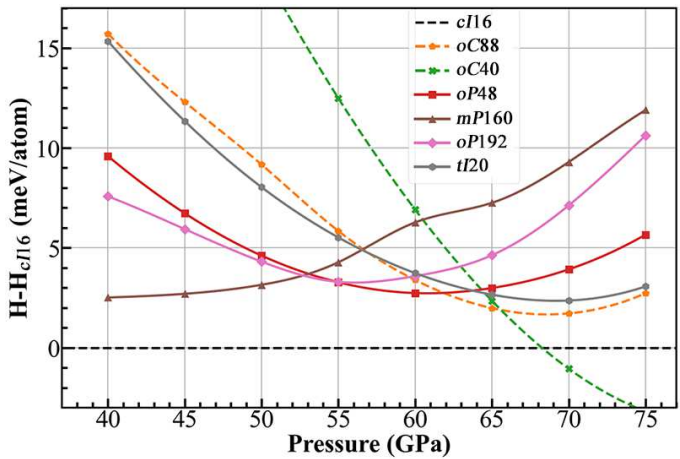}\\[5pt]  
\caption{DFT enthalpy per atom for experimental (dashed line) and predicted structures (solid line) of Li as a function of pressure between 40 and 75 GPa relative to the $cI16$ structure.
}
\label{fig:Fig2}
\end{center}
\end{figure}

The $mP160$ phase [Fig. \ref{fig:Fig1}(a)] has the $P1$ space group symmetry with 160 atoms in the unit cell and atomic chains arranged in parallel along the $c$ axis. Every four chains form a rhombic prism stacked alternatively in the {\it b-c} plane. Several similar structures including the recently proposed dimerized $C2/m$ structure \cite{Wang2022} also have been predicted by our searches. These structures have smaller unit cells and higher symmetries [see Supplemental Material~\cite{Note1}, Sec.~SIIIA], and they are different from the $mP160$ in the arrangement of the prisms and are nearly degenerate with the $mP160$ phase at 50 and 55 GPa, but become unfavorable at higher or lower pressures. An interesting phenomenon observed in the $mP160$ structure is that the atomic chains are gradually distorted with increasing pressure, similar to structural changes during melting of a crystal (Fig. S5). This phenomenon is not observed in other structures of this type due to the constraints of the higher symmetry and smaller unit cells, and it is responsible for the favorable enthalpy of the $mP160$ phase of Li. Another notable phenomenon observed in the $mP160$ phase, as well as other structures described below, is the wide distribution of distances of the nearest-neighbor contacts (Fig. S5). At 50 GPa, the Li...Li distances are between 2.01$-$2.47 \AA. This bond alternation is also found in the experimentally identified $oC88$ and $oC40$ phases, and is a common feature of Li at high pressure due to Peierls distortion \cite{Lv2011,Yao.2009.10.1103/physrevlett.102.115503,Neaton.1999.10.1038/22067}.

The $oP192$ phase [Fig.\ref{fig:Fig1}(b)] in the $Pcc2$ symmetry contains 192 atoms in the unit cell. It is a $mP160$-like structure but formed by more distorted rhombic prisms (along the $c$ axis), with wrinkled atomic layers in {\it a-b} plane stacked along the $c$ axis. 

The $oP48$ phase [Fig.\ref{fig:Fig1}(c)] shares similar structural features with the $oC88$ and $oC40$ structures, exhibiting a 3D irregular network [see top of Fig. \ref{fig:Fig1}(c)] and containing a series of atomic layers [see bottom of Fig. \ref{fig:Fig1}(c)]. The $oP48$ phase adopts the $Pbcn$ symmetry, with six crystallographically distinct Li atoms in the eightfold $8b$ Wyckoff site, forming two symmetry-related 8-atom layers (shown in grey) and two symmetry-related 16-atom layers (shown in black). 

The $tI20$ phase [Fig.\ref{fig:Fig1}(d)] adopts the $I4$ symmetry in a commensurate host-guest structure. The host structure comprises double helixes along the $c$ axis (shown in black), while the guest structure is composed of Li-Li dimers located at the body center of the tetragonal cell (shown in grey). One of the most intriguing discoveries in alkalis under pressure is the emergence of incommensurate host-guest structures in Na-V, K-III and Rb-IV \cite{Gregoryanz2008,McMahon.2006.10.1103/physrevb.74.140102,McMahon.2001.10.1103/physrevlett.87.055501,McMahon.2006.10.1039/b517777b}. The $tI20$ phase of Li is a counterpart of this type of structures. 

We have performed enthalpy calculations using DFT at a high level of accuracy for the four newly predicted Li structures along with those of the experimentally observed $cI16$, $oC88$, and $oC40$ phases, and the results are plotted as a function of pressure in Fig.~\ref{fig:Fig2}. Our results for the known Li structures are in good agreement with previous theoretical calculations \cite{Marques2011,Gorelli.2012.10.1103/physrevlett.108.055501}. The $cI16$ phase is stable up to $\sim$67 GPa, beyond which the semiconducting $oC40$ phase is enthalpically favorable; meanwhile, the $oC88$ phase is less favorable than the $cI16$ phase at all pressures (about 2 meV higher in enthalpy than the $cI16$ at 65 GPa). Experimentally, the $oC88$ phase was observed at 77 K above $\sim$62 GPa \cite{Guillaume2011}, and a theoretical calculation assessed its Gibbs free energy under harmonic approximation and found that the free energy of the $oC88$ phase at 65 GPa was lower than that of the $cI16$ phase at finite temperatures \cite{Gorelli.2012.10.1103/physrevlett.108.055501}. Similarly, all four predicted structures of Li are less favorable in enthalpy than the $cI16$ phase by a few meV. However, these structures show lower enthalpies than the $oC88$ phase between $40-62$ GPa, and among them the $mP160$ phase has the lowest enthalpy below 53 GPa, while the $oP192$ and $oP48$ phases have the lowest enthalpies at 55 and 60 GPa, respectively. The $tI20$ phase is nearly degenerate with the $oC88$ phase in the entire pressure range considered. 

The dynamic stability of the $oP48$ and $tI20$ phases was verified by the absence of imaginary frequency in the Brillouin zone via phonon calculations (Supplemental Material \cite{Note1} Sec. SI, Fig. S1). For the $mP160$ and $oP192$ phases, the large unit cells make DFT-based phonon calculations prohibitively expensive, thus the dynamic stability is examined by Deep Potential molecular dynamics (DPMD) simulations at $T = 100$ K, using the $NPT$ ensemble, at 45 and 55 GPa, respectively. The mean-square displacements remain nearly constant during the 1 $ns$ simulation period, indicating that these two structures are dynamically stable (Supplemental Material \cite{Note1}, Sec. IIIB, Fig. S6).

\begin{figure}
\begin{center}
\includegraphics[width=0.50\textwidth]{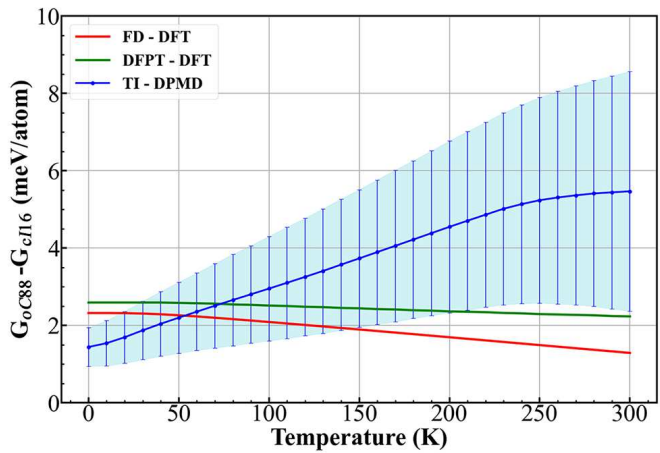}\\[5pt] 
\caption{Gibbs free energy of the $oC88$ structure as a function of temperature relative to the $cI16$ structure at 65 GPa, calculated under the harmonic approximation using both FD and DFPT methods through DFT, and considering anharmonicity using TI through DPMD. Error bars are shown for the TI results, which are estimated from the DP error and statistical uncertainty in the MD simulations.}
\label{fig:Fig3}
\end{center}
\end{figure}

The thermodynamic stability of various Li phases near the melting minimum were determined by their relative Gibbs free energy. We first revisited the relative stability between the experimentally observed $cI16$ and $oC88$ phases by calculating the Gibbs free energy via two approaches: (i) under harmonic approximation using both finite displacement (FD) and density functional perturbation theory (DFPT) methods within DFT (see Supplemental Material \cite{Note1}, Sec.~SI), and (ii) considering anharmonicity using thermodynamic integration (TI) through DPMD (see Supplemental Material \cite{Note1}, Sec.~SIV). The DP error of the Li-DP-Hyb2 model relative to DFT increases with increasing temperature, but stays less than 1~meV/atom at temperatures below 150 K (see Supplemental Material \cite{Note1}, Sec. SIIC).  The statistical uncertainty for molecular dynamics (MD) calculations was estimated to be well below 1 meV/atom. The calculated Gibbs free energy as a function of temperature is shown in Fig.~\ref{fig:Fig3}. All the calculations produced higher free energy of the $oC88$ phase than that of the $cI16$ phase. Under harmonic approximation, the FD and DFPT methods give consistent results with a difference of less than 1 meV/atom. The relative free energy of the $oC88$ phase slightly decreases with increasing temperature, but remains above that of the $cI16$ phase up to 300 K. At 0 K, the zero-point energies for the $cI16$ and $oC88$ phases calculated by FD (DFPT) are 83.1 (83.0) and 83.2 (83.4) meV/atom, respectively. This result implies that while NQEs contribute significantly to the free energy, they are largely canceled between the two structures, in agreement with previous \textit{ab initio} path-integral MD studies~\cite{Feng2015,Wang2022}. Considering anharmonicity, there is a different trend of relative free energy of the $oC88$ phase, which increases rapidly with increasing temperature. This result indicates that anharmonic effects tend to destabilize the $oC88$ phase and play an increasingly important role at increasing temperatures. Overall, our present results strongly suggest metastability of the $oC88$ phase against the $cI16$ phase and highlight a major role of anharmonic effects on determining the relative stability among Li phases.

\begin{figure}
\begin{center}
\includegraphics[width=0.52\textwidth]{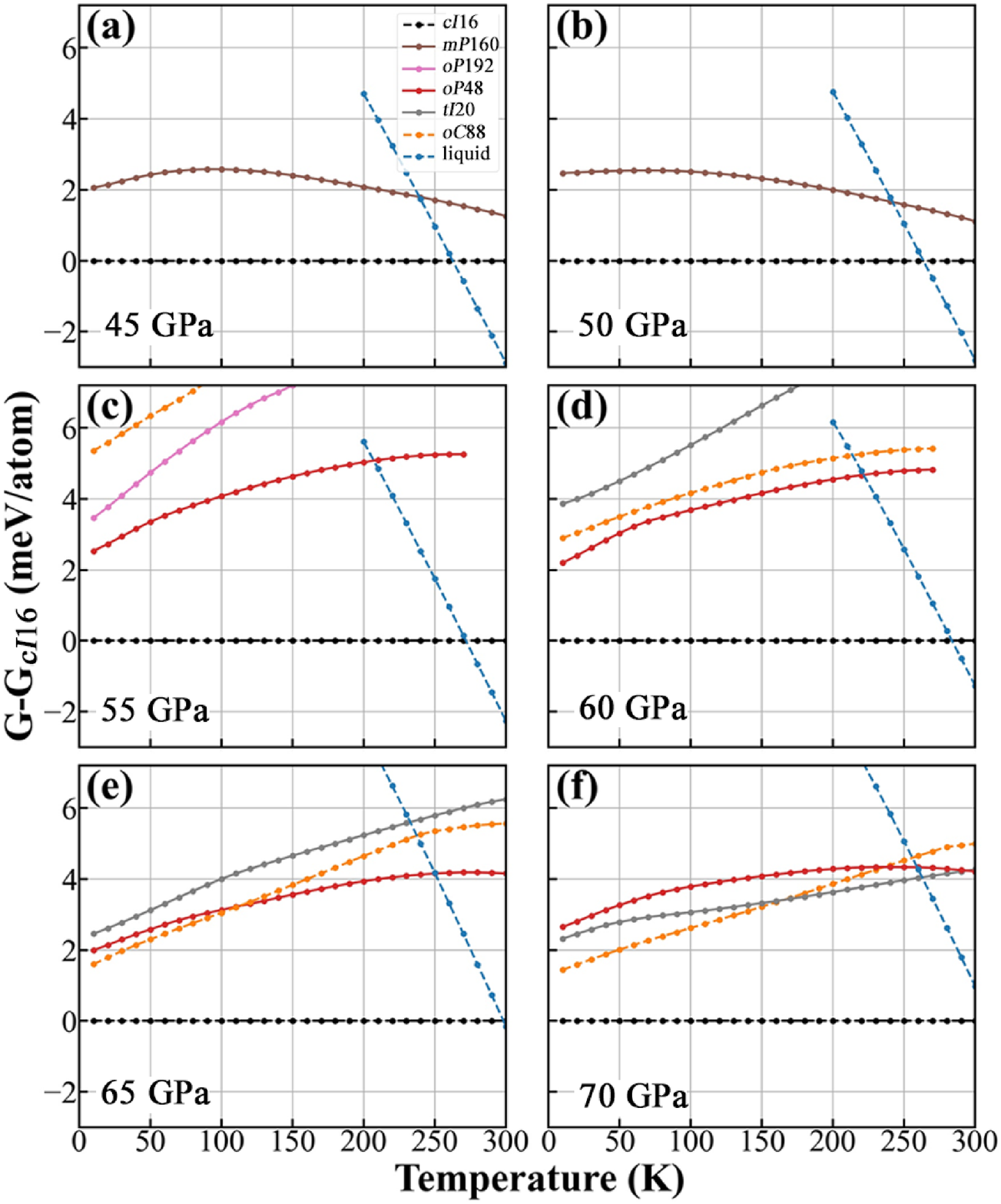}\\[5pt]  
\caption{Gibbs free energies of various Li structures relative to the $cI16$ phase as a function of temperature, calculated using TI through DPMD at (a) 45, (b) 50, (c) 55, (d) 60, (e) 65, and (f) 70 GPa.
}
\label{fig:Fig4}
\end{center}
\end{figure}

We now proceed to assess the thermodynamic stability of the four predicted Li structures by calculating their Gibbs free energy using TI through DPMD. The free energies calculated at 45, 50, 55, 60, 65, and 70 GPa of the predicted, experimental and liquid phases of Li are plotted as function of temperature up to 300 K are shown in Fig. \ref{fig:Fig4}. The results show that the melting temperature is 262 K at 45 GPa and gradually increases to above 300 K at 70 GPa. 

At 45 and 50 GPa [Fig. \ref{fig:Fig4}(a) and (b)], only the $cI16$ and $mP160$ phases are dynamically stable, and the $cI16$ phase is lower in free energy by only $\sim$2-4 meV/atom. Note that the relative free energy of the $mP160$ phase gradually decreases with increasing temperature at 45 and 50 GPa. This leads to free energy differences among the $cI16$, $mP160$ and liquid phases lower than 2 meV at 250 K. At 55 and 60 GPa [Fig. \ref{fig:Fig4}(c) and (d)], the predicted $oP48$ phase becomes the second most stable structure until melting. The predicted $oP192$ phase is dynamically stable only at 55 GPa and shows lower free energies than the $oC88$ phase. At 65 and 70 GPa [pressures where the $oC88$ phase is observed in experiment, Fig. \ref{fig:Fig4}(e) and (f)], the $oC88$ phase is the second most stable structure at low temperatures, while the $oP48$ and $tI20$ phases possess lower free energies than the $oC88$ phase above $\sim$110 K (at 65 GPa) and above $\sim$160 K (at 70 GPa), respectively. 
The small differences in free energy between various Li phases are sometimes within the errors of the current DPMD simulations, highlighting the difficulty in theoretically assessing the relative stability of Li phases around the melting minimum. However, several insights can be gained from the present work: (1) The $mP160$ phase is the most promising candidate structure at finite temperatures below 50 GPa, as it exhibits nearly degenerated Gibbs free energy with the $cI16$ and liquid phases at 50 GPa and 250 K, where all other competing phases are dynamically unstable; (2) the $oP48$ phase exhibits superior thermodynamic stability among all the predicted structures and the experimentally observed $oC88$ phase at 55 GPa; (3) Li exhibits a nearly flat energy landscape with multiple shallow local minima around the melting minimum, on which the predicted structures compete with the previously reported $cI16$ and $oC88$ phases within a small free energy window of only a few meV. These considerations suggest that the four newly predicted Li structures should be experimentally accessible, and their occurrence would be sensitive to experimental P-T paths.

In summary, we demonstrate in this work that the combination of the state-of-the-art swarm-intelligence-based global optimization and deep leaning techniques allows to explore and construct the PES of materials in unprecedented scales and details. By taking the challenging problem of Li phases around the melting minimum as a prominent case study, we discovered four complex crystal structures with large unit cells containing up to 192 atoms that host competing Gibbs free energies comparable with those previously observed in experiments. A recent experimental work observed X-ray diffraction peaks of crystalline phases of Li in this P-T region through a rapid compression scheme by taking advantage of high X-ray flux at modern synchrotron \cite{Frost2019}. However, the data are not sufficient to unambiguously assign structures due to the difficulties associated with the sample containment and the degradation of the diamond cell. The present work offers strong evidence for the existence of new solid phases of dense Li around the melting minimum, and our findings are expected to stimulate further theoretical and experimental exploration of this intriguing problem.

\begin{acknowledgments}
J.L. grateful to Prof. G. Csanyi for valuable discussions on construction of machine learning potentials. The work of J.L. is supported by National Natural Science Foundation of China (Grants No. 12034009, 91961204, 11974134 and 11904129). 
The work of H.W. is supported by the National Science Foundation of China (Grant No.11871110 and 12122103).
\end{acknowledgments}


\begin{thebibliography}{35}%
\makeatletter
\providecommand \@ifxundefined [1]{%
 \@ifx{#1\undefined}
}%
\providecommand \@ifnum [1]{%
 \ifnum #1\expandafter \@firstoftwo
 \else \expandafter \@secondoftwo
 \fi
}%
\providecommand \@ifx [1]{%
 \ifx #1\expandafter \@firstoftwo
 \else \expandafter \@secondoftwo
 \fi
}%
\providecommand \natexlab [1]{#1}%
\providecommand \enquote  [1]{``#1''}%
\providecommand \bibnamefont  [1]{#1}%
\providecommand \bibfnamefont [1]{#1}%
\providecommand \citenamefont [1]{#1}%
\providecommand \href@noop [0]{\@secondoftwo}%
\providecommand \href [0]{\begingroup \@sanitize@url \@href}%
\providecommand \@href[1]{\@@startlink{#1}\@@href}%
\providecommand \@@href[1]{\endgroup#1\@@endlink}%
\providecommand \@sanitize@url [0]{\catcode `\\12\catcode `\$12\catcode
  `\&12\catcode `\#12\catcode `\^12\catcode `\_12\catcode `\%12\relax}%
\providecommand \@@startlink[1]{}%
\providecommand \@@endlink[0]{}%
\providecommand \url  [0]{\begingroup\@sanitize@url \@url }%
\providecommand \@url [1]{\endgroup\@href {#1}{\urlprefix }}%
\providecommand \urlprefix  [0]{URL }%
\providecommand \Eprint [0]{\href }%
\providecommand \doibase [0]{https://doi.org/}%
\providecommand \selectlanguage [0]{\@gobble}%
\providecommand \bibinfo  [0]{\@secondoftwo}%
\providecommand \bibfield  [0]{\@secondoftwo}%
\providecommand \translation [1]{[#1]}%
\providecommand \BibitemOpen [0]{}%
\providecommand \bibitemStop [0]{}%
\providecommand \bibitemNoStop [0]{.\EOS\space}%
\providecommand \EOS [0]{\spacefactor3000\relax}%
\providecommand \BibitemShut  [1]{\csname bibitem#1\endcsname}%
\let\auto@bib@innerbib\@empty
\bibitem [{\citenamefont {Shimizu}\ \emph {et~al.}(2002)\citenamefont
  {Shimizu}, \citenamefont {Ishikawa}, \citenamefont {Takao}, \citenamefont
  {Yagi},\ and\ \citenamefont {Amaya}}]{Shimizu2002}%
  \BibitemOpen
  \bibfield  {author} {\bibinfo {author} {\bibfnamefont {K.}~\bibnamefont
  {Shimizu}}, \bibinfo {author} {\bibfnamefont {H.}~\bibnamefont {Ishikawa}},
  \bibinfo {author} {\bibfnamefont {D.}~\bibnamefont {Takao}}, \bibinfo
  {author} {\bibfnamefont {T.}~\bibnamefont {Yagi}},\ and\ \bibinfo {author}
  {\bibfnamefont {K.}~\bibnamefont {Amaya}},\ }\href
  {https://doi.org/10.1038/nature01098} {\bibfield  {journal} {\bibinfo
  {journal} {Nature}\ }\textbf {\bibinfo {volume} {419}},\ \bibinfo {pages}
  {597} (\bibinfo {year} {2002})}\BibitemShut {NoStop}%
\bibitem [{\citenamefont {Struzhkin}\ \emph {et~al.}(2002)\citenamefont
  {Struzhkin}, \citenamefont {Eremets}, \citenamefont {Gan}, \citenamefont
  {Mao},\ and\ \citenamefont
  {Hemley}}]{Struzhkin.2002.10.1126/science.1078535}%
  \BibitemOpen
  \bibfield  {author} {\bibinfo {author} {\bibfnamefont {V.~V.}\ \bibnamefont
  {Struzhkin}}, \bibinfo {author} {\bibfnamefont {M.~I.}\ \bibnamefont
  {Eremets}}, \bibinfo {author} {\bibfnamefont {W.}~\bibnamefont {Gan}},
  \bibinfo {author} {\bibfnamefont {H.-k.}\ \bibnamefont {Mao}},\ and\ \bibinfo
  {author} {\bibfnamefont {R.~J.}\ \bibnamefont {Hemley}},\ }\href
  {https://doi.org/10.1126/science.1078535} {\bibfield  {journal} {\bibinfo
  {journal} {Science}\ }\textbf {\bibinfo {volume} {298}},\ \bibinfo {pages}
  {1213} (\bibinfo {year} {2002})}\BibitemShut {NoStop}%
\bibitem [{\citenamefont {Ma}\ \emph {et~al.}(2009)\citenamefont {Ma},
  \citenamefont {Eremets}, \citenamefont {Oganov}, \citenamefont {Xie},
  \citenamefont {Trojan}, \citenamefont {Medvedev}, \citenamefont {Lyakhov},
  \citenamefont {Valle},\ and\ \citenamefont {Prakapenka}}]{Ma2009}%
  \BibitemOpen
  \bibfield  {author} {\bibinfo {author} {\bibfnamefont {Y.}~\bibnamefont
  {Ma}}, \bibinfo {author} {\bibfnamefont {M.}~\bibnamefont {Eremets}},
  \bibinfo {author} {\bibfnamefont {A.~R.}\ \bibnamefont {Oganov}}, \bibinfo
  {author} {\bibfnamefont {Y.}~\bibnamefont {Xie}}, \bibinfo {author}
  {\bibfnamefont {I.}~\bibnamefont {Trojan}}, \bibinfo {author} {\bibfnamefont
  {S.}~\bibnamefont {Medvedev}}, \bibinfo {author} {\bibfnamefont {A.~O.}\
  \bibnamefont {Lyakhov}}, \bibinfo {author} {\bibfnamefont {M.}~\bibnamefont
  {Valle}},\ and\ \bibinfo {author} {\bibfnamefont {V.}~\bibnamefont
  {Prakapenka}},\ }\href {https://doi.org/10.1038/nature07786} {\bibfield
  {journal} {\bibinfo  {journal} {Nature}\ }\textbf {\bibinfo {volume} {458}},\
  \bibinfo {pages} {182} (\bibinfo {year} {2009})}\BibitemShut {NoStop}%
\bibitem [{\citenamefont {Matsuoka}\ and\ \citenamefont
  {Shimizu}(2009)}]{Matsuoka2009}%
  \BibitemOpen
  \bibfield  {author} {\bibinfo {author} {\bibfnamefont {T.}~\bibnamefont
  {Matsuoka}}\ and\ \bibinfo {author} {\bibfnamefont {K.}~\bibnamefont
  {Shimizu}},\ }\href {https://doi.org/10.1038/nature07827} {\bibfield
  {journal} {\bibinfo  {journal} {Nature}\ }\textbf {\bibinfo {volume} {458}},\
  \bibinfo {pages} {186} (\bibinfo {year} {2009})}\BibitemShut {NoStop}%
\bibitem [{\citenamefont {Lv}\ \emph {et~al.}(2011)\citenamefont {Lv},
  \citenamefont {Wang}, \citenamefont {Zhu},\ and\ \citenamefont
  {Ma}}]{Lv2011}%
  \BibitemOpen
  \bibfield  {author} {\bibinfo {author} {\bibfnamefont {J.}~\bibnamefont
  {Lv}}, \bibinfo {author} {\bibfnamefont {Y.}~\bibnamefont {Wang}}, \bibinfo
  {author} {\bibfnamefont {L.}~\bibnamefont {Zhu}},\ and\ \bibinfo {author}
  {\bibfnamefont {Y.}~\bibnamefont {Ma}},\ }\href
  {https://doi.org/10.1103/PhysRevLett.106.015503} {\bibfield  {journal}
  {\bibinfo  {journal} {Phys. Rev. Lett.}\ }\textbf {\bibinfo {volume}
  {106}},\ \bibinfo {pages} {015503} (\bibinfo {year} {2011})}\BibitemShut
  {NoStop}%
\bibitem [{\citenamefont {Marqu{\'{e}}s}\ \emph {et~al.}(2011)\citenamefont
  {Marqu{\'{e}}s}, \citenamefont {McMahon}, \citenamefont {Gregoryanz},
  \citenamefont {Hanfland}, \citenamefont {Guillaume}, \citenamefont {Pickard},
  \citenamefont {Ackland},\ and\ \citenamefont {Nelmes}}]{Marques2011}%
  \BibitemOpen
  \bibfield  {author} {\bibinfo {author} {\bibfnamefont {M.}~\bibnamefont
  {Marqu{\'{e}}s}}, \bibinfo {author} {\bibfnamefont {M.~I.}\ \bibnamefont
  {McMahon}}, \bibinfo {author} {\bibfnamefont {E.}~\bibnamefont {Gregoryanz}},
  \bibinfo {author} {\bibfnamefont {M.}~\bibnamefont {Hanfland}}, \bibinfo
  {author} {\bibfnamefont {C.~L.}\ \bibnamefont {Guillaume}}, \bibinfo {author}
  {\bibfnamefont {C.~J.}\ \bibnamefont {Pickard}}, \bibinfo {author}
  {\bibfnamefont {G.~J.}\ \bibnamefont {Ackland}},\ and\ \bibinfo {author}
  {\bibfnamefont {R.~J.}\ \bibnamefont {Nelmes}},\ }\href
  {https://doi.org/10.1103/PhysRevLett.106.095502} {\bibfield  {journal}
  {\bibinfo  {journal} {Phys. Rev. Lett.}\ }\textbf {\bibinfo {volume}
  {106}},\ \bibinfo {pages} {095502} (\bibinfo {year} {2011})}\BibitemShut
  {NoStop}%
\bibitem [{\citenamefont {Gregoryanz}\ \emph {et~al.}(2005)\citenamefont
  {Gregoryanz}, \citenamefont {Degtyareva}, \citenamefont {Somayazulu},
  \citenamefont {Hemley},\ and\ \citenamefont {Mao}}]{Gregoryanz2005}%
  \BibitemOpen
  \bibfield  {author} {\bibinfo {author} {\bibfnamefont {E.}~\bibnamefont
  {Gregoryanz}}, \bibinfo {author} {\bibfnamefont {O.}~\bibnamefont
  {Degtyareva}}, \bibinfo {author} {\bibfnamefont {M.}~\bibnamefont
  {Somayazulu}}, \bibinfo {author} {\bibfnamefont {R.~J.}\ \bibnamefont
  {Hemley}},\ and\ \bibinfo {author} {\bibfnamefont {H.-k.}\ \bibnamefont
  {Mao}},\ }\href {https://doi.org/10.1103/PhysRevLett.94.185502} {\bibfield
  {journal} {\bibinfo  {journal} {Phys. Rev. Lett.}\ }\textbf {\bibinfo
  {volume} {94}},\ \bibinfo {pages} {185502} (\bibinfo {year}
  {2005})}\BibitemShut {NoStop}%
\bibitem [{\citenamefont {Guillaume}\ \emph {et~al.}(2011)\citenamefont
  {Guillaume}, \citenamefont {Gregoryanz}, \citenamefont {Degtyareva},
  \citenamefont {McMahon}, \citenamefont {Hanfland}, \citenamefont {Evans},
  \citenamefont {Guthrie}, \citenamefont {Sinogeikin},\ and\ \citenamefont
  {Mao}}]{Guillaume2011}%
  \BibitemOpen
  \bibfield  {author} {\bibinfo {author} {\bibfnamefont {C.~L.}\ \bibnamefont
  {Guillaume}}, \bibinfo {author} {\bibfnamefont {E.}~\bibnamefont
  {Gregoryanz}}, \bibinfo {author} {\bibfnamefont {O.}~\bibnamefont
  {Degtyareva}}, \bibinfo {author} {\bibfnamefont {M.~I.}\ \bibnamefont
  {McMahon}}, \bibinfo {author} {\bibfnamefont {M.}~\bibnamefont {Hanfland}},
  \bibinfo {author} {\bibfnamefont {S.}~\bibnamefont {Evans}}, \bibinfo
  {author} {\bibfnamefont {M.}~\bibnamefont {Guthrie}}, \bibinfo {author}
  {\bibfnamefont {S.~V.}\ \bibnamefont {Sinogeikin}},\ and\ \bibinfo {author}
  {\bibfnamefont {H.}~\bibnamefont {Mao}},\ }\href
  {https://doi.org/10.1038/nphys1864} {\bibfield  {journal} {\bibinfo
  {journal} {Nat. Phys.}\ }\textbf {\bibinfo {volume} {7}},\ \bibinfo
  {pages} {211} (\bibinfo {year} {2011})}\BibitemShut {NoStop}%
\bibitem [{\citenamefont {Schaeffer}\ \emph {et~al.}(2012)\citenamefont
  {Schaeffer}, \citenamefont {Talmadge}, \citenamefont {Temple},\ and\
  \citenamefont {Deemyad}}]{Schaeffer2012}%
  \BibitemOpen
  \bibfield  {author} {\bibinfo {author} {\bibfnamefont {A.~M.~J.}\
  \bibnamefont {Schaeffer}}, \bibinfo {author} {\bibfnamefont {W.~B.}\
  \bibnamefont {Talmadge}}, \bibinfo {author} {\bibfnamefont {S.~R.}\
  \bibnamefont {Temple}},\ and\ \bibinfo {author} {\bibfnamefont
  {S.}~\bibnamefont {Deemyad}},\ }\href
  {https://doi.org/10.1103/PhysRevLett.109.185702} {\bibfield  {journal}
  {\bibinfo  {journal} {Phys. Rev. Lett.}\ }\textbf {\bibinfo {volume}
  {109}},\ \bibinfo {pages} {185702} (\bibinfo {year} {2012})}\BibitemShut
  {NoStop}%
\bibitem [{\citenamefont {Frost}\ \emph {et~al.}(2019)\citenamefont {Frost},
  \citenamefont {Kim}, \citenamefont {McBride}, \citenamefont {Peterson},
  \citenamefont {Smith}, \citenamefont {Sun},\ and\ \citenamefont
  {Glenzer}}]{Frost2019}%
  \BibitemOpen
  \bibfield  {author} {\bibinfo {author} {\bibfnamefont {M.}~\bibnamefont
  {Frost}}, \bibinfo {author} {\bibfnamefont {J.~B.}\ \bibnamefont {Kim}},
  \bibinfo {author} {\bibfnamefont {E.~E.}\ \bibnamefont {McBride}}, \bibinfo
  {author} {\bibfnamefont {J.~R.}\ \bibnamefont {Peterson}}, \bibinfo {author}
  {\bibfnamefont {J.~S.}\ \bibnamefont {Smith}}, \bibinfo {author}
  {\bibfnamefont {P.}~\bibnamefont {Sun}},\ and\ \bibinfo {author}
  {\bibfnamefont {S.~H.}\ \bibnamefont {Glenzer}},\ }\href
  {https://doi.org/10.1103/PhysRevLett.123.065701} {\bibfield  {journal}
  {\bibinfo  {journal} {Phys. Rev. Lett.}\ }\textbf {\bibinfo {volume}
  {123}},\ \bibinfo {pages} {065701} (\bibinfo {year} {2019})}\BibitemShut
  {NoStop}%
\bibitem [{\citenamefont {Hanfland}\ \emph {et~al.}(2000)\citenamefont
  {Hanfland}, \citenamefont {Syassen}, \citenamefont {Christensen},\ and\
  \citenamefont {Novikov}}]{Hanfland2000}%
  \BibitemOpen
  \bibfield  {author} {\bibinfo {author} {\bibfnamefont {M.}~\bibnamefont
  {Hanfland}}, \bibinfo {author} {\bibfnamefont {K.}~\bibnamefont {Syassen}},
  \bibinfo {author} {\bibfnamefont {N.~E.}\ \bibnamefont {Christensen}},\ and\
  \bibinfo {author} {\bibfnamefont {D.~L.}\ \bibnamefont {Novikov}},\ }\href
  {https://doi.org/10.1038/35041515} {\bibfield  {journal} {\bibinfo  {journal}
  {Nature}\ }\textbf {\bibinfo {volume} {408}},\ \bibinfo {pages} {174}
  (\bibinfo {year} {2000})}\BibitemShut {NoStop}%
\bibitem [{\citenamefont {Gregoryanz}\ \emph {et~al.}(2008)\citenamefont
  {Gregoryanz}, \citenamefont {Lundegaard}, \citenamefont {McMahon},
  \citenamefont {Guillaume}, \citenamefont {Nelmes},\ and\ \citenamefont
  {Mezouar}}]{Gregoryanz2008}%
  \BibitemOpen
  \bibfield  {author} {\bibinfo {author} {\bibfnamefont {E.}~\bibnamefont
  {Gregoryanz}}, \bibinfo {author} {\bibfnamefont {L.~F.}\ \bibnamefont
  {Lundegaard}}, \bibinfo {author} {\bibfnamefont {M.~I.}\ \bibnamefont
  {McMahon}}, \bibinfo {author} {\bibfnamefont {C.}~\bibnamefont {Guillaume}},
  \bibinfo {author} {\bibfnamefont {R.~J.}\ \bibnamefont {Nelmes}},\ and\
  \bibinfo {author} {\bibfnamefont {M.}~\bibnamefont {Mezouar}},\ }\href
  {https://doi.org/10.1126/science.1155715} {\bibfield  {journal} {\bibinfo
  {journal} {Science}\ }\textbf {\bibinfo {volume} {320}},\ \bibinfo {pages}
  {1054} (\bibinfo {year} {2008})}\BibitemShut {NoStop}%
\bibitem [{\citenamefont {Ackland}\ \emph {et~al.}(2017)\citenamefont
  {Ackland}, \citenamefont {Dunuwille}, \citenamefont {Martinez-Canales},
  \citenamefont {Loa}, \citenamefont {Zhang}, \citenamefont {Sinogeikin},
  \citenamefont {Cai},\ and\ \citenamefont {Deemyad}}]{Ackland2017}%
  \BibitemOpen
  \bibfield  {author} {\bibinfo {author} {\bibfnamefont {G.~J.}\ \bibnamefont
  {Ackland}}, \bibinfo {author} {\bibfnamefont {M.}~\bibnamefont {Dunuwille}},
  \bibinfo {author} {\bibfnamefont {M.}~\bibnamefont {Martinez-Canales}},
  \bibinfo {author} {\bibfnamefont {I.}~\bibnamefont {Loa}}, \bibinfo {author}
  {\bibfnamefont {R.}~\bibnamefont {Zhang}}, \bibinfo {author} {\bibfnamefont
  {S.}~\bibnamefont {Sinogeikin}}, \bibinfo {author} {\bibfnamefont
  {W.}~\bibnamefont {Cai}},\ and\ \bibinfo {author} {\bibfnamefont
  {S.}~\bibnamefont {Deemyad}},\ }\href
  {https://doi.org/10.1126/science.aal4886} {\bibfield  {journal} {\bibinfo
  {journal} {Science}\ }\textbf {\bibinfo {volume} {356}},\ \bibinfo {pages}
  {1254} (\bibinfo {year} {2017})}\BibitemShut {NoStop}%
\bibitem [{\citenamefont {Wang}\ \emph
  {et~al.}(2022{\natexlab{a}})\citenamefont {Wang}, \citenamefont {Wang},
  \citenamefont {Hermann}, \citenamefont {Pan}, \citenamefont {Shi},
  \citenamefont {Wang}, \citenamefont {Xing},\ and\ \citenamefont
  {Sun}}]{Wang2022}%
  \BibitemOpen
  \bibfield  {author} {\bibinfo {author} {\bibfnamefont {Y.}~\bibnamefont
  {Wang}}, \bibinfo {author} {\bibfnamefont {J.}~\bibnamefont {Wang}}, \bibinfo
  {author} {\bibfnamefont {A.}~\bibnamefont {Hermann}}, \bibinfo {author}
  {\bibfnamefont {S.}~\bibnamefont {Pan}}, \bibinfo {author} {\bibfnamefont
  {J.}~\bibnamefont {Shi}}, \bibinfo {author} {\bibfnamefont {H.-T.}\
  \bibnamefont {Wang}}, \bibinfo {author} {\bibfnamefont {D.}~\bibnamefont
  {Xing}},\ and\ \bibinfo {author} {\bibfnamefont {J.}~\bibnamefont {Sun}},\
  }\href {https://doi.org/10.1103/PhysRevB.105.214101} {\bibfield  {journal}
  {\bibinfo  {journal} {Phys. Rev. B}\ }\textbf {\bibinfo {volume}
  {105}},\ \bibinfo {pages} {214101} (\bibinfo {year}
  {2022}{\natexlab{a}})}\BibitemShut {NoStop}%
\bibitem [{\citenamefont {Luedemann}\ and\ \citenamefont
  {Kennedy}(1968)}]{Luedemann.1968.10.1029/jb073i008p02795}%
  \BibitemOpen
  \bibfield  {author} {\bibinfo {author} {\bibfnamefont {H.~D.}\ \bibnamefont
  {Luedemann}}\ and\ \bibinfo {author} {\bibfnamefont {G.~C.}\ \bibnamefont
  {Kennedy}},\ }\href {https://doi.org/10.1029/jb073i008p02795} {\bibfield
  {journal} {\bibinfo  {journal} {J. Geophys. Res.}\ }\textbf
  {\bibinfo {volume} {73}},\ \bibinfo {pages} {2795} (\bibinfo {year}
  {1968})}\BibitemShut {NoStop}%
\bibitem [{\citenamefont {McBride}\ \emph {et~al.}(2015)\citenamefont
  {McBride}, \citenamefont {Munro}, \citenamefont {Stinton}, \citenamefont
  {Husband}, \citenamefont {Briggs}, \citenamefont {Liermann},\ and\
  \citenamefont {McMahon}}]{McBride.2015.10.1103/physrevb.91.144111}%
  \BibitemOpen
  \bibfield  {author} {\bibinfo {author} {\bibfnamefont {E.~E.}\ \bibnamefont
  {McBride}}, \bibinfo {author} {\bibfnamefont {K.~A.}\ \bibnamefont {Munro}},
  \bibinfo {author} {\bibfnamefont {G.~W.}\ \bibnamefont {Stinton}}, \bibinfo
  {author} {\bibfnamefont {R.~J.}\ \bibnamefont {Husband}}, \bibinfo {author}
  {\bibfnamefont {R.}~\bibnamefont {Briggs}}, \bibinfo {author} {\bibfnamefont
  {H.-P.}\ \bibnamefont {Liermann}},\ and\ \bibinfo {author} {\bibfnamefont
  {M.~I.}\ \bibnamefont {McMahon}},\ }\href
  {https://doi.org/10.1103/physrevb.91.144111} {\bibfield  {journal} {\bibinfo
  {journal} {Phys. Rev. B}\ }\textbf {\bibinfo {volume} {91}},\ \bibinfo
  {pages} {144111} (\bibinfo {year} {2015})}\BibitemShut {NoStop}%
\bibitem [{\citenamefont {Robinson}\ \emph {et~al.}(2019)\citenamefont
  {Robinson}, \citenamefont {Zong}, \citenamefont {Ackland}, \citenamefont
  {Woolman},\ and\ \citenamefont
  {Hermann}}]{Robinson.2019.10.1073/pnas.1900985116}%
  \BibitemOpen
  \bibfield  {author} {\bibinfo {author} {\bibfnamefont {V.~N.}\ \bibnamefont
  {Robinson}}, \bibinfo {author} {\bibfnamefont {H.}~\bibnamefont {Zong}},
  \bibinfo {author} {\bibfnamefont {G.~J.}\ \bibnamefont {Ackland}}, \bibinfo
  {author} {\bibfnamefont {G.}~\bibnamefont {Woolman}},\ and\ \bibinfo {author}
  {\bibfnamefont {A.}~\bibnamefont {Hermann}},\ }\href
  {https://doi.org/10.1073/pnas.1900985116} {\bibfield  {journal} {\bibinfo
  {journal} {PNAS}\ }\textbf
  {\bibinfo {volume} {116}},\ \bibinfo {pages} {201900985} (\bibinfo {year}
  {2019})}\BibitemShut {NoStop}%
\bibitem [{\citenamefont {Tamblyn}\ \emph {et~al.}(2008)\citenamefont
  {Tamblyn}, \citenamefont {Raty},\ and\ \citenamefont {Bonev}}]{Tamblyn2008}%
  \BibitemOpen
  \bibfield  {author} {\bibinfo {author} {\bibfnamefont {I.}~\bibnamefont
  {Tamblyn}}, \bibinfo {author} {\bibfnamefont {J.-Y.}\ \bibnamefont {Raty}},\
  and\ \bibinfo {author} {\bibfnamefont {S.~A.}\ \bibnamefont {Bonev}},\ }\href
  {https://doi.org/10.1103/PhysRevLett.101.075703} {\bibfield  {journal}
  {\bibinfo  {journal} {Phys. Rev. Lett.}\ }\textbf {\bibinfo {volume}
  {101}},\ \bibinfo {pages} {075703} (\bibinfo {year} {2008})}\BibitemShut
  {NoStop}%
\bibitem [{\citenamefont {Hern{\'{a}}ndez}\ \emph {et~al.}(2010)\citenamefont
  {Hern{\'{a}}ndez}, \citenamefont {Rodriguez-Prieto}, \citenamefont
  {Bergara},\ and\ \citenamefont {Alf{\`{e}}}}]{Hernandez2010}%
  \BibitemOpen
  \bibfield  {author} {\bibinfo {author} {\bibfnamefont {E.~R.}\ \bibnamefont
  {Hern{\'{a}}ndez}}, \bibinfo {author} {\bibfnamefont {A.}~\bibnamefont
  {Rodriguez-Prieto}}, \bibinfo {author} {\bibfnamefont {A.}~\bibnamefont
  {Bergara}},\ and\ \bibinfo {author} {\bibfnamefont {D.}~\bibnamefont
  {Alf{\`{e}}}},\ }\href {https://doi.org/10.1103/PhysRevLett.104.185701}
  {\bibfield  {journal} {\bibinfo  {journal} {Phys. Rev. Lett.}\
  }\textbf {\bibinfo {volume} {104}},\ \bibinfo {pages} {185701} (\bibinfo
  {year} {2010})}\BibitemShut {NoStop}%
\bibitem [{\citenamefont {Feng}\ \emph {et~al.}(2015)\citenamefont {Feng},
  \citenamefont {Chen}, \citenamefont {Alf{\`{e}}}, \citenamefont {Li},\ and\
  \citenamefont {Wang}}]{Feng2015}%
  \BibitemOpen
  \bibfield  {author} {\bibinfo {author} {\bibfnamefont {Y.}~\bibnamefont
  {Feng}}, \bibinfo {author} {\bibfnamefont {J.}~\bibnamefont {Chen}}, \bibinfo
  {author} {\bibfnamefont {D.}~\bibnamefont {Alf{\`{e}}}}, \bibinfo {author}
  {\bibfnamefont {X.-Z.}\ \bibnamefont {Li}},\ and\ \bibinfo {author}
  {\bibfnamefont {E.}~\bibnamefont {Wang}},\ }\href
  {https://doi.org/10.1063/1.4907752} {\bibfield  {journal} {\bibinfo
  {journal} {J. Chem. Phys.}\ }\textbf {\bibinfo {volume}
  {142}},\ \bibinfo {pages} {064506} (\bibinfo {year} {2015})}\BibitemShut
  {NoStop}%
\bibitem [{\citenamefont {Elatresh}\ \emph {et~al.}(2016)\citenamefont
  {Elatresh}, \citenamefont {Bonev}, \citenamefont {Gregoryanz},\ and\
  \citenamefont {Ashcroft}}]{Elatresh2016}%
  \BibitemOpen
  \bibfield  {author} {\bibinfo {author} {\bibfnamefont {S.~F.}\ \bibnamefont
  {Elatresh}}, \bibinfo {author} {\bibfnamefont {S.~A.}\ \bibnamefont {Bonev}},
  \bibinfo {author} {\bibfnamefont {E.}~\bibnamefont {Gregoryanz}},\ and\
  \bibinfo {author} {\bibfnamefont {N.~W.}\ \bibnamefont {Ashcroft}},\ }\href
  {https://doi.org/10.1103/PhysRevB.94.104107} {\bibfield  {journal} {\bibinfo
  {journal} {Phys. Rev. B}\ }\textbf {\bibinfo {volume} {94}},\ \bibinfo
  {pages} {104107} (\bibinfo {year} {2016})}\BibitemShut {NoStop}%
\bibitem [{\citenamefont {Wang}\ \emph {et~al.}(2010)\citenamefont {Wang},
  \citenamefont {Lv}, \citenamefont {Zhu},\ and\ \citenamefont
  {Ma}}]{Wang2010}%
  \BibitemOpen
  \bibfield  {author} {\bibinfo {author} {\bibfnamefont {Y.}~\bibnamefont
  {Wang}}, \bibinfo {author} {\bibfnamefont {J.}~\bibnamefont {Lv}}, \bibinfo
  {author} {\bibfnamefont {L.}~\bibnamefont {Zhu}},\ and\ \bibinfo {author}
  {\bibfnamefont {Y.}~\bibnamefont {Ma}},\ }\href
  {https://doi.org/10.1103/PhysRevB.82.094116} {\bibfield  {journal} {\bibinfo
  {journal} {Phys. Rev. B}\ }\textbf {\bibinfo {volume} {82}},\ \bibinfo
  {pages} {094116} (\bibinfo {year} {2010})}\BibitemShut {NoStop}%
\bibitem [{\citenamefont {Wang}\ \emph {et~al.}(2012)\citenamefont {Wang},
  \citenamefont {Lv}, \citenamefont {Zhu},\ and\ \citenamefont
  {Ma}}]{Wang2012}%
  \BibitemOpen
  \bibfield  {author} {\bibinfo {author} {\bibfnamefont {Y.}~\bibnamefont
  {Wang}}, \bibinfo {author} {\bibfnamefont {J.}~\bibnamefont {Lv}}, \bibinfo
  {author} {\bibfnamefont {L.}~\bibnamefont {Zhu}},\ and\ \bibinfo {author}
  {\bibfnamefont {Y.}~\bibnamefont {Ma}},\ }\href
  {https://doi.org/10.1016/j.cpc.2012.05.008} {\bibfield  {journal} {\bibinfo
  {journal} {Comput. Phys. Commun.}\ }\textbf {\bibinfo {volume}
  {183}},\ \bibinfo {pages} {2063} (\bibinfo {year} {2012})}\BibitemShut
  {NoStop}%
\bibitem [{\citenamefont {Lv}\ \emph {et~al.}(2012)\citenamefont {Lv},
  \citenamefont {Wang}, \citenamefont {Zhu},\ and\ \citenamefont
  {Ma}}]{Lv2012}%
  \BibitemOpen
  \bibfield  {author} {\bibinfo {author} {\bibfnamefont {J.}~\bibnamefont
  {Lv}}, \bibinfo {author} {\bibfnamefont {Y.}~\bibnamefont {Wang}}, \bibinfo
  {author} {\bibfnamefont {L.}~\bibnamefont {Zhu}},\ and\ \bibinfo {author}
  {\bibfnamefont {Y.}~\bibnamefont {Ma}},\ }\href
  {https://doi.org/10.1063/1.4746757} {\bibfield  {journal} {\bibinfo
  {journal} {J. Chem. Phys.}\ }\textbf {\bibinfo {volume}
  {137}},\ \bibinfo {pages} {084104} (\bibinfo {year} {2012})}\BibitemShut
  {NoStop}%
\bibitem [{\citenamefont {Gao}\ \emph {et~al.}(2019)\citenamefont {Gao},
  \citenamefont {Gao}, \citenamefont {Lu}, \citenamefont {Lv}, \citenamefont
  {Wang},\ and\ \citenamefont {Ma}}]{Gao2019}%
  \BibitemOpen
  \bibfield  {author} {\bibinfo {author} {\bibfnamefont {B.}~\bibnamefont
  {Gao}}, \bibinfo {author} {\bibfnamefont {P.}~\bibnamefont {Gao}}, \bibinfo
  {author} {\bibfnamefont {S.}~\bibnamefont {Lu}}, \bibinfo {author}
  {\bibfnamefont {J.}~\bibnamefont {Lv}}, \bibinfo {author} {\bibfnamefont
  {Y.}~\bibnamefont {Wang}},\ and\ \bibinfo {author} {\bibfnamefont
  {Y.}~\bibnamefont {Ma}},\ }\href {https://doi.org/10.1016/j.scib.2019.02.009}
  {\bibfield  {journal} {\bibinfo  {journal} {Sci. Bull.}\ }\textbf
  {\bibinfo {volume} {64}},\ \bibinfo {pages} {301} (\bibinfo {year}
  {2019})}\BibitemShut {NoStop}%
\bibitem [{\citenamefont {Zhang}\ \emph {et~al.}(2018)\citenamefont {Zhang},
  \citenamefont {Han}, \citenamefont {Wang}, \citenamefont {Car},\ and\
  \citenamefont {Weinan}}]{Zhang2018}%
  \BibitemOpen
  \bibfield  {author} {\bibinfo {author} {\bibfnamefont {L.}~\bibnamefont
  {Zhang}}, \bibinfo {author} {\bibfnamefont {J.}~\bibnamefont {Han}}, \bibinfo
  {author} {\bibfnamefont {H.}~\bibnamefont {Wang}}, \bibinfo {author}
  {\bibfnamefont {R.}~\bibnamefont {Car}},\ and\ \bibinfo {author}
  {\bibfnamefont {E.}~\bibnamefont {Weinan}},\ }\href
  {https://doi.org/10.1103/PhysRevLett.120.143001} {\bibfield  {journal}
  {\bibinfo  {journal} {Phys. Rev. Lett.}\ }\textbf {\bibinfo {volume}
  {120}},\ \bibinfo {pages} {143001} (\bibinfo {year} {2018})}\BibitemShut
  {NoStop}%
\bibitem [{\citenamefont {Han}\ \emph {et~al.}(2018)\citenamefont {Han},
  \citenamefont {Zhang}, \citenamefont {Car},\ and\ \citenamefont
  {E}}]{Han2018}%
  \BibitemOpen
  \bibfield  {author} {\bibinfo {author} {\bibfnamefont {J.}~\bibnamefont
  {Han}}, \bibinfo {author} {\bibfnamefont {L.}~\bibnamefont {Zhang}}, \bibinfo
  {author} {\bibfnamefont {R.}~\bibnamefont {Car}},\ and\ \bibinfo {author}
  {\bibfnamefont {W.}~\bibnamefont {E}},\ }\href
  {https://doi.org/10.4208/cicp.oa-2017-0213} {\bibfield  {journal} {\bibinfo
  {journal} {Commun. Comput. Phys.}\ }\textbf {\bibinfo
  {volume} {23}},\ \bibinfo {pages} {629} (\bibinfo {year} {2018})}\BibitemShut
  {NoStop}%
\bibitem [{\citenamefont {Wang}\ \emph
  {et~al.}(2022{\natexlab{b}})\citenamefont {Wang}, \citenamefont {Lv},
  \citenamefont {Gao},\ and\ \citenamefont {Ma}}]{Wang2022Crystal}%
  \BibitemOpen
  \bibfield  {author} {\bibinfo {author} {\bibfnamefont {Y.}~\bibnamefont
  {Wang}}, \bibinfo {author} {\bibfnamefont {J.}~\bibnamefont {Lv}}, \bibinfo
  {author} {\bibfnamefont {P.}~\bibnamefont {Gao}},\ and\ \bibinfo {author}
  {\bibfnamefont {Y.}~\bibnamefont {Ma}},\ }\href
  {https://doi.org/10.1021/acs.accounts.2c00243} {\bibfield  {journal}
  {\bibinfo  {journal} {Acc. Chem. Res.}\ }\textbf {\bibinfo
  {volume} {55}},\ \bibinfo {pages} {2068} (\bibinfo {year}
  {2022}{\natexlab{b}})}\BibitemShut {NoStop}%
\bibitem [{Not()}]{Note1}%
  \BibitemOpen
  \href@noop {} {}\bibinfo {note} {See Supplemental Material at
  http://link.aps.org/XXXX for detailed information on the method, simulation
  protocols, additional results and structure information}\BibitemShut
  {NoStop}%
\bibitem [{\citenamefont {Yao}\ \emph {et~al.}(2009)\citenamefont {Yao},
  \citenamefont {Tse},\ and\ \citenamefont
  {Klug}}]{Yao.2009.10.1103/physrevlett.102.115503}%
  \BibitemOpen
  \bibfield  {author} {\bibinfo {author} {\bibfnamefont {Y.}~\bibnamefont
  {Yao}}, \bibinfo {author} {\bibfnamefont {J.~S.}\ \bibnamefont {Tse}},\ and\
  \bibinfo {author} {\bibfnamefont {D.~D.}\ \bibnamefont {Klug}},\ }\href
  {https://doi.org/10.1103/physrevlett.102.115503} {\bibfield  {journal}
  {\bibinfo  {journal} {Phys. Rev. Lett.}\ }\textbf {\bibinfo {volume}
  {102}},\ \bibinfo {pages} {115503} (\bibinfo {year} {2009})}\BibitemShut
  {NoStop}%
\bibitem [{\citenamefont {Neaton}\ and\ \citenamefont
  {Ashcroft}(1999)}]{Neaton.1999.10.1038/22067}%
  \BibitemOpen
  \bibfield  {author} {\bibinfo {author} {\bibfnamefont {J.~B.}\ \bibnamefont
  {Neaton}}\ and\ \bibinfo {author} {\bibfnamefont {N.~W.}\ \bibnamefont
  {Ashcroft}},\ }\href {https://doi.org/10.1038/22067} {\bibfield  {journal}
  {\bibinfo  {journal} {Nature}\ }\textbf {\bibinfo {volume} {400}},\ \bibinfo
  {pages} {141} (\bibinfo {year} {1999})}\BibitemShut {NoStop}%
\bibitem [{\citenamefont {McMahon}\ \emph {et~al.}(2006)\citenamefont
  {McMahon}, \citenamefont {Nelmes}, \citenamefont {Schwarz},\ and\
  \citenamefont {Syassen}}]{McMahon.2006.10.1103/physrevb.74.140102}%
  \BibitemOpen
  \bibfield  {author} {\bibinfo {author} {\bibfnamefont {M.~I.}\ \bibnamefont
  {McMahon}}, \bibinfo {author} {\bibfnamefont {R.~J.}\ \bibnamefont {Nelmes}},
  \bibinfo {author} {\bibfnamefont {U.}~\bibnamefont {Schwarz}},\ and\ \bibinfo
  {author} {\bibfnamefont {K.}~\bibnamefont {Syassen}},\ }\href
  {https://doi.org/10.1103/physrevb.74.140102} {\bibfield  {journal} {\bibinfo
  {journal} {Phys. Rev. B}\ }\textbf {\bibinfo {volume} {74}},\ \bibinfo
  {pages} {140102} (\bibinfo {year} {2006})}\BibitemShut {NoStop}%
\bibitem [{\citenamefont {McMahon}\ \emph {et~al.}(2001)\citenamefont
  {McMahon}, \citenamefont {Rekhi},\ and\ \citenamefont
  {Nelmes}}]{McMahon.2001.10.1103/physrevlett.87.055501}%
  \BibitemOpen
  \bibfield  {author} {\bibinfo {author} {\bibfnamefont {M.~I.}\ \bibnamefont
  {McMahon}}, \bibinfo {author} {\bibfnamefont {S.}~\bibnamefont {Rekhi}},\
  and\ \bibinfo {author} {\bibfnamefont {R.~J.}\ \bibnamefont {Nelmes}},\
  }\href {https://doi.org/10.1103/physrevlett.87.055501} {\bibfield  {journal}
  {\bibinfo  {journal} {Phys. Rev. Lett.}\ }\textbf {\bibinfo {volume}
  {87}},\ \bibinfo {pages} {055501} (\bibinfo {year} {2001})}\BibitemShut
  {NoStop}%
\bibitem [{\citenamefont {McMahon}\ and\ \citenamefont
  {Nelmes}(2006)}]{McMahon.2006.10.1039/b517777b}%
  \BibitemOpen
  \bibfield  {author} {\bibinfo {author} {\bibfnamefont {M.~I.}\ \bibnamefont
  {McMahon}}\ and\ \bibinfo {author} {\bibfnamefont {R.~J.}\ \bibnamefont
  {Nelmes}},\ }\href {https://doi.org/10.1039/b517777b} {\bibfield  {journal}
  {\bibinfo  {journal} {Chem. Soc. Rev.}\ }\textbf {\bibinfo {volume}
  {35}},\ \bibinfo {pages} {943} (\bibinfo {year} {2006})}\BibitemShut
  {NoStop}%
\bibitem [{\citenamefont {Gorelli}\ \emph {et~al.}(2012)\citenamefont
  {Gorelli}, \citenamefont {Elatresh}, \citenamefont {Guillaume}, \citenamefont
  {Marqu{\'{e}}s}, \citenamefont {Ackland}, \citenamefont {Santoro}, \citenamefont
  {Bonev},\ and\ \citenamefont
  {Gregoryanz}}]{Gorelli.2012.10.1103/physrevlett.108.055501}%
  \BibitemOpen
  \bibfield  {author} {\bibinfo {author} {\bibfnamefont {F.~A.}\ \bibnamefont
  {Gorelli}}, \bibinfo {author} {\bibfnamefont {S.~F.}\ \bibnamefont
  {Elatresh}}, \bibinfo {author} {\bibfnamefont {C.~L.}\ \bibnamefont
  {Guillaume}}, \bibinfo {author} {\bibfnamefont {M.}~\bibnamefont {Marqu{\'{e}}s}},
  \bibinfo {author} {\bibfnamefont {G.~J.}\ \bibnamefont {Ackland}}, \bibinfo
  {author} {\bibfnamefont {M.}~\bibnamefont {Santoro}}, \bibinfo {author}
  {\bibfnamefont {S.~A.}\ \bibnamefont {Bonev}},\ and\ \bibinfo {author}
  {\bibfnamefont {E.}~\bibnamefont {Gregoryanz}},\ }\href
  {https://doi.org/10.1103/physrevlett.108.055501} {\bibfield  {journal}
  {\bibinfo  {journal} {Phys. Rev. Lett.}\ }\textbf {\bibinfo {volume}
  {108}},\ \bibinfo {pages} {055501} (\bibinfo {year} {2012})}\BibitemShut
  {NoStop}%
\end{thebibliography}
\bibliographystyle{apsrev4-2}

\end{document}